\documentclass[12pt]{emulateapj}
\usepackage{longtable}
\usepackage[fleqn]{amsmath}
\usepackage{amsthm}
\usepackage{color,ulem}

\def\eg{{\it e.g.\,}}
\def\msun{\;{\rm M_\odot}}
\def\mbh{{\rm M_\bullet}}
\def\cf{{\it c.f.\,}}
\def\ie{{\it i.e.\,}}

\shorttitle{Rapidly Reorienting AGN Jets in Galaxy Cluster Cores}
\begin{document}

\title{Isotropic Heating of Galaxy Cluster Cores via Rapidly Reorienting AGN Jets}

\author{Arif Babul}
\affil{Department of Physics and Astronomy, University of Victoria, Victoria, BC, V8P 1A1, Canada}
\email{babul@uvic.ca}

\author{Prateek Sharma}
\affil{Department of Physics and Joint Astronomy Program, Indian Institute of Science, Bangalore 560012, India}

\and

\author{Christopher S. Reynolds}
\affil{Department of Astronomy, University of Maryland, College Park, MD 20742-2421, USA}

%% Notice that some of these authors has alternate affiliations, which


\begin{abstract}
AGN jets carry more than sufficient energy to stave off catastrophic cooling of the intracluster medium (ICM) in the cores of cool-core clusters.   However, in order to prevent catastrophic cooling, the ICM must be heated in a near-isotropic fashion and narrow bipolar jets with $P_{\rm jet}=10^{44-45}$ ergs/s, typical of radio AGNs at cluster centres, are inefficient at heating the gas in the transverse direction to the jets.   We argue that due to existent conditions in cluster cores, the SMBHs will, in addition to accreting gas via radiatively inefficient flows, experience short stochastic episodes of enhanced accretion via thin discs.  In general, the orientation of these accretion discs will be misaligned with the spin axis of the black holes and the ensuing torques will cause the black hole's spin axis (and therefore, the jet axis) to slew and rapidly change direction.   This model not only explains recent observations showing successive generations of jet-lobes-bubbles in individual cool-core clusters that are offset from each other in the angular direction with respect to the cluster center, but also shows that AGN jets {\it can} heat the cluster core nearly isotropically on the gas cooling timescale.     Our model {\it does} require that the  SMBHs at the centers of cool-core clusters be spinning relatively slowly.  Torques from individual misaligned discs are ineffective at tilting rapidly spinning black holes by more than a few degrees.  Additionally,  since SMBHs that host thin accretion discs will manifest as quasars, we predict that roughly 1--2 rich clusters within $z<0.5$ should have quasars at their centers. 
\end{abstract}


\section{Introduction}\label{sec-intro}

The cool-core conundrum poses a critical challenge for theoretical models 
seeking to explain the observed properties of clusters of galaxies.  
The intracluster medium (ICM) 
in cool-core clusters should, according to simple theoretical arguments and direct 
observations of X-ray emission~\citep{fabian84}, be cooling  and dropping out 
at prodigious rates, yet only relatively meager amounts of cold gas is seen in cluster centers. 
High resolution  X-ray and radio observations of the cores of these clusters indicate that the dominant central
galaxy (hereafter referred to as the BCG --- the brightest cluster galaxy)
invariably shows evidence of active galactic nuclei (AGN) behaviour, often 
in the form of powerful bipolar jets and pairs of approximately spherical depressions 
in the X-ray emissions, typically interpreted as being due to bubbles of relativistic plasma
that have been inflated in the ICM by the jets.   The inferred power of the jets is 
comparable to the radiative losses in the ICM  and the case --- based on theoretical 
arguments and observational inference --- in favour of this radio-mode AGN energy injection 
into the ICM being the most likely explanation for the diminished cooling, 
is fairly robust \citep{mcn07}.   However, the 
precise manner in which this energy is injected into and distributed within the ICM 
remains an open question.   The mechanics of how an apparently narrow bipolar outflow is able to 
heat the ICM in a near-isotropic fashion is particularly vexing\citep[\cf][]{vr06}.   

 In a recent simulation study, \citet{gaspari11} claim to have solved 
the ``isotropy'' problem in cool-core galaxy clusters and successfully stave off catastrophic cooling over a 
cosmological timescale.  In their most successful and best explored scheme, the jets are modelled as 
explosive, massive, subrelativistic outflows with $P_{\rm jet}=10^{47-48}$ ergs/s and a duty cycle of  
$\sim 6\%$.   The nearly isotropic heating of the cool core is affected by local shocks and strong turbulence 
induced by the powerful, explosive outbursts \citep[see also][]{gas11}.  Success, however, comes at a cost. 
The mass accretion rate 
(onto the black hole) required to power the jets is $\dot{{\rm M}}\approx 0.8 \left({P_{\rm jet}/ 10^{47}\;
{\rm ergs/s}}\right) \dot{{\rm M}}_{\rm Edd}$, where $\dot{{\rm M}}_{\rm Edd}\equiv ({L_{\rm Edd}/ 
{0.1c^2}}) \approx 22\;\mbh_{9}\;\msun\;{\rm yr}^{-1}$ is the Eddington accretion rate and  $\mbh_{9}
\equiv (\mbh/10^9\msun)$ is the black hole mass in units of $10^9\msun$.  Apart from the jets, an AGN 
accreting at such high rates will also radiate copiously and will be identified as a quasar.   Observations; 
however, do not support this.   Neither H1821+643 nor B0910+410, the only two AGNs at the centres of 
cool-core clusters in the $z<0.5$ universe that are quasars, show any evidence of current jet activity 
\citep{rus10,osullivan2012}.   While the AGN in M87 though actively driving a jet, is not 
a quasar and neither was the recent burst of jet activity by the central AGN in Perseus \citep{nagai10} 
accompanied by quasar-like emissions.

Recent detailed observational studies of individual cool core groups and clusters suggest an alternate solution to the 
``isotropy'' problem.  Many of the systems show evidence of multiple generations of radio jets and cocoons as well 
as  X-ray cavity pairs 
\citep[\cf][]{mcnamara2001,dunn06,wise2007,forman07,gia2011,randall2011,osullivan2012}.  In a number of these systems 
(\eg~M87 [Virgo], NGC 1275 [Perseus], NGC 4636, NGC 5044, NGC 5813, Hydra A, MACS J0913.7+4056 [also known as 
CL09104+4109 and hereafter referred to as CL09]), successive 
generations of jet-lobe-bubbles are significantly offset from each other in the 
angular direction on the sky, with respect to the cluster center.   If the AGN jet byproducts can trace a nearly 
isotropic angular distribution about the cluster center, then so should the associated heating.
And if the timescale over which the jet  energy is redistributed is shorter than the cooling time, 
the isotropy problem is a non-issue.   In the discussion to follow, we assert that ``isotropic heating'' and the angular 
misalignment of jet byproducts are related processes. The question then is:  What is the most likely explanation for the  
the observations given the conditions in the cores of  cool-core clusters?  

Potential explanations can be classified into two broad categories:  The first invoke interactions 
between the jet and the ``ICM weather'', \ie large-scale\footnote{We use the label ``large-scale'' to 
distinguish between velocity fluctuations on scales of a few kiloparsecs and larger,
as opposed to ``small-scale'' turbulence generated by, for example, supernova blastwaves.} turbulence, wakes and 
bulk velocities induced either by mergers and  orbiting substructure \citep{heinz06,morsony10}.  Undoubtedly, 
jet-ICM interactions must be occurring at some level; however, there are 
a number of factors that suggest that its impact  in cores of real cool-core clusters is less than in the simulations. 
For example, the simulation  study by \citet{morsony10} shows that strong large-scale turbulent flows triggered 
by mergers {\it can} result in large angular displacement of the low density jet material (\eg~bubbles) from the jet axis;  
however,  the required flow velocities are present only in unrelaxed cluster cores; \ie, cores that are either being
churned by an ongoing major merger or are in the recovery phase following a major merger.  Most cool-core 
clusters do not show any evidence of disturbances associated with recent major mergers \citep{pooleII}.   In fact,  comparative 
analyses of X-ray and lensing cluster data (\eg~\citealt{mahdavi08,mahdavi12,zhang10}) indicate that cool cluster cores 
are dynamically relaxed.   Additionally, the simulations show that large-scale merger-induced turbulence has no impact 
on the direction of the high velocity jet flow on small scales and therefore, cannot account for the apparent change of 
orientation in the jet directions on subkiloparsec to kiloparsec scales as observed in, for example, CL09, Virgo and Perseus
(\cf~\S\ref{sec-observe}, Table 1).

The alternate class of models invoke changes in the orientation of the spin axis of the supermassive black holes 
(SMBHs) that are powering the jets to explain the misalignment between the successive generations of jets-lobes-bubbles.   
This explanation is premised on the understanding that (a) the angular position of any particular feature (jet/cavity/bubble) 
is indicative of the direction of the jet axis at the time when the feature was formed, and (b) the  jet axis is always 
coincident with the black hole's spin axis.    The spin orientation of black holes can change as a result of  precession 
\citep[\cf][and references therein]{piz05b,lodato06,liu07,cam07a}, spin flips \citep[\cf][and references therein]{mer02,cam07b,kesden10} or  slewing (or tilting) of the black hole spin axis \citep[\cf][]{sch96,kin05,lodato06}.    We will consider these options in more detail 
in \S\ref{sec-observe}, after we review some of the most compelling observations of  repeated re-orientation of the jet 
axes in galaxy groups and clusters.   For now, we will simply note that the only two  viable scenarios are: (a) precession 
associated with binary black hole systems in the cluster cores, or (b) stochastic slewing of the black hole spin axis due to 
torques from recurring, short-lived, misaligned thin discs.  In this paper, we argue that the observations are best understood 
in the context of the latter model.

In the next section, we will first review the most compelling observational evidence for recurring re-orientation of the jet axes in cluster environments. We then assess the various scenarios involving changes in the direction of the black hole's spin axis and identify those that are both plausible and compatible with the observations.   In \S\ref{sec-model}, we motivate the case for the our preferred model, which involves rapid, stochastic reorientations of the SMBH's spin axis on relatively short timescales, summarize the associated physics, and discuss why we expect this to be a recurrent phenomenon in cores of cool-core clusters.  In \S\ref{sec-discuss}, we discuss the astrophysical implications of our model.

\section{Wandering Jet Axis:\break Assessing the Observational Evidence}\label{sec-observe}

\begin{deluxetable*}{p{3.5 cm}ccclp{3.5 cm}}
\tabletypesize{\scriptsize}
\tablecaption{Features Indicating Jet Reorientation in Perseus and Virgo \label{tbl-1}}
\tablewidth{17 cm}
\tablehead{
\colhead{Feature} & \colhead{P.A. (North)\tablenotemark{a}} & \colhead{P.A. (South)\tablenotemark{a}}  &
\colhead{Timescale} & \colhead{Ref.} & \colhead{Comment}}
\startdata
\multicolumn{2}{l}{{\bf Perseus (NGC 1275, 3C 84)}}  & & &   \\
\\
Milliarcsecond jet & --- & $\sim170^{\rm o}$ & --- & 1,2  & 
\citet{kri93} argue that the jet is nearly aligned with line of sight (los)
very close to the core.  On kpc scales, however, it is 
$\sim 45^{\rm o}$ relative to los, implying a significant change in direction.\\[+3 mm]
Cocoon-like feature at $r \sim 1.2$ mas & --- & $\sim210^{\rm o}$ & --- & 1,2,3 & \\[+3 mm]
Parsec-scale cocoon at $r\sim 12$ mas & $\sim {5^{\rm o}}-10^{{\rm o}}$ & $\sim 180^{{\rm o}}$ & $\sim 25$ yrs & 1,3,4 & \\[+3 mm]
Inner bubble & $\sim {345^{\rm o}}$ & $\sim 155^{{\rm o}}$ & $\sim10$ Myrs & 5 & Note that, unlike here, angles in \citet{dunn06} were measured in the clockwise direction.
\\[+3 mm]
Outer bubble &  $\sim {5^{\rm o}}$ & $\sim 215^{{\rm o}}$ & $\sim 15-20$ Myrs  & 5 & \\[+3 mm]
Ghost bubble &  $\sim {305^{\rm o}}$ & $\sim 170^{{\rm o}}$ & $\sim 70-75$ Myrs  & 5 &
Based on the shape of the northern bubble and the kinematics of the trailing H$\alpha$ filaments, \citet{dunn06} argue that the bubble must be moving nearly perpendicular to the line of sight --- in the plane of the sky.\\[+3 mm]
Ancient bubble &  $\sim {345^{\rm o}}$ & $\sim 230^{{\rm o}}$ & $\sim 100$ Myrs & 5 & \\
 \\
 \multicolumn{2}{l}{{\bf Virgo (M87, 3C 274)}}  & & &   \\
 \\
Jet & ${290^{\rm o}}$ & ${115^{\rm o}}$  & --- & 6, 7 & \citet{biretta95} argue that from 0.1 to 1000 parsecs, the observed jet (stretching to the northwest) is aligned at  $\sim 40^{\rm o}$ to los.  The counter jet is not observed but its direction is inferred from the location of the 
hot spot.\\[+3 mm]
Inner lobes (in radio); jet/counter-jet cavities (in X-ray) at $\sim 2.5$kpc & ${270^{\rm o}-275^{\rm o}}$ & ${120^{\rm o}-130^{\rm o}}$  & --- & 6, 7, 8 & \\[+3 mm]
Series of buoyant bubbles  in the southeast  & --- &  ${160^{\rm o}, 140^{\rm o}, 120^{\rm o}, 100^{\rm o}}$ & $\langle\Delta t_{\rm age}\rangle\sim 6$ Myrs & 8 & Ordered from closest to furthest in projection \\[+3 mm]
Intermediate East-West jets and lobes & ${270^{\rm o}}$ &   ${90^{\rm o}}$ &   $\sim 20$ Myrs  & 9, 10 & Following \citet{klein99}, we identity the Eastern radio arm  (Feature C in Fig.3 of \citet{owen00}), terminating in the ``ear-like Feature B, and the Western extension of the jet along with ``knot D'' as jet-lobe features.  \citet{klein99} refers to Features B and D as ``intermediate lobes".\\[+3 mm]
Outer X-ray cavity & ${35^{\rm o}-40^{\rm o}}$ &   & $70$ Myrs & 8 &  This is the cavity discovered by \citet{forman07} and named ``outer cavity". \\[+3 mm]
Outer radio bubbles & ${10^{\rm o}}$ & ${205^{\rm o}}$ & $100$ Myrs & 9, 10 & \\
\enddata
\tablenotetext{a}{P.A. is the angle in the sky at the source location between North and the feature, measured in the counterclockwise direction.}
\tablenotetext{\ }{(1)\citet{dha98}, (2) \citet{lister01}, (3) \citet{kri93}, (4) \citet{nagai09}, (5) \citet{dunn06}, (6) \citet{biretta93}, (7) \citet{biretta95}, (8) \citet{forman07}, (9)\citet{owen00}, (10)\citet{klein99}}
\end{deluxetable*}

Observations of  ``X-shaped"  radio galaxies (XRGs) \citep[\cf][and references therein]{hk10}, double-double radio galaxies (DDRGs)  \citep[\cf][and references therein]{joshi11},
and changes in the orientation of the jets on sub-kiloparsec scales in Seyfert galaxies \citep[\cf][and references therein]{gall06} have long hinted at the possibility that AGN jets undergo discreet changes in direction by  moderately large angles on relatively short  (a few Myrs to several 10s of Myrs) timescales.   However, recent radio and X-ray studies of AGN activity in cool-core groups and clusters offer further and arguably more compelling evidence for  ``jet reorientation''.   This evidence is manifest in a variety of forms ranging from large-angle bends in the radio jets, to abrupt changes in direction from one jet episode to the next, to multiple generations of radio cocoons and X-ray cavities, some of which appear to trace out a nearly isotropic distribution in projection about the cluster center.   Examples of such systems include CL09, M87 [Virgo], NGC 1275 [Perseus], NGC 4636, NGC 5044, NGC 5813, and Hydra A \citep{klein99,dunn06,forman07,wise2007,gia2011,randall2011,osullivan2012}.    

In Table \ref{tbl-1}, we focus on the Virgo and the Perseus clusters, two systems that have been the subject of  a detailed, multi-wavelength observational campaign over the course of the past decade, and catalogue most of the pertinent features seen in the radio and X-ray maps of these two systems typically associated with the radio-mode AGN activity of the central SMBH.  Collectively, these observations of misaligned active and relic jets, as well as radio lobes, bubbles, and distinct X-ray cavities separated by large angles, strongly indicate that the underlying outflow responsible for forming these features changes direction  between energetic jet ``episodes''   over timescales ranging from few Myrs to few 10s of Myrs.   (Fig. 1 in \citealt{dunn06} and Fig. 5 in \citealt{forman07} show this visually.)     On the sub-parsec and parsec 
scales, this is indicated by changes in the direction of the jets themselves, and on kiloparsec scales and beyond, several generations of radio bubbles/X-ray cavities are observed and their distribution covers almost all projected angles, with many cavity pairs orientated nearly collinear with the SMBH.   These relative changes in the orientation of the jet-counterjet features are more consistent with rotation of the jet axis rather than displacement  of the source, which would tend to generate wide-angle tail-like geometries \citep[\eg][and references therein]{jetha08}. 

The most straightforward way to affect a change in the direction of the BH's spin axis is via accretion.   Supermassive black holes in BCGs at the centers of cool core clusters are thought to be accreting gas from their surroundings  via  geometrically thick, radiatively inefficient flows at a relatively low rate $\dot{{\rm M}} < 10^{-3} \dot{{\rm M}}_{\rm Edd}$ \citep{all06}.   In the event that the angular momentum of the geometrically thick flow is misaligned with the direction of the BH's spin axis, the latter will gradually swivel and move to align with the former as BH accretes the gas and its angular momentum.  The corresponding alignment timescale, however,  is comparable to the black hole's mass-doubling timescale ($\sim \mbh/\dot{{\rm M}}$):  For a $10^9\msun$ SMBH accreting at a rate $\dot{{\rm M}}\approx 0.1 \msun {\rm yr}^{-1}$,  the corresponding timescale is $\sim 10$ Gyrs.   Accretion via geometrically thick flow is not a viable option. 

 The fastest way to change the  direction of an AGN's jet axis is via spin-flip, an abrupt change in the direction of a black hole's spin axis following a BH-BH merger.   This mechanism is one of the leading explanation for the XRGs and misaligned DDRGs \citep{mez12,mar12}.  Spin-flips, however, are unlikely to explain the Virgo and Perseus observations, for instance, because one would need to invoke several BH-BH mergers over a duration of $\sim$100 Myrs to account for the $5$--$7$ episodes of changes in the direction of the jet axes over this period, which is  implausible.  

A number of authors, including \citet{klein99,piz05b,dunn06,falceta10} have invoked  precession of the black hole spin axis about a fixed axis to explain the distribution of radio and X-ray observations in Virgo and Perseus.   To account for the observations, the precession of the BH must be combined with a jet model in which the outflows are intermittent and of short duration (compared to the precession period) so that the outcome is a sequence of discreet misaligned cavities.  Precession can occur if the black hole in question is part of a binary black hole pair whose spins and the orbital angular momentum are misaligned.  The precession timescale, when both black holes are of comparable mass, is \citep{mer05,key11} 

\begin{equation}
t_{\rm BH, prec}\sim 2.4 \times 10^7\; {\rm yrs}  \left({{\cal{A}}\over 1\;{\rm pc}}\right)^{5/2} {\mbh_9}^{-3/2},
\end{equation}
where ${\cal{A}}$ is the binary semi-major axis.  This timescale is reasonable and unless the  binary SMBH are embedded in a massive, gaseous accretion disc, the lifetime of the binary ought to be more than long enough to span multiple precession cycles  (\citealt{mer05}), which is what \citet{dunn06} require in order to account for the cavities in Perseus. 
Moreover,  it is not inconceivable that the BCGs in cluster cores host binary supermassive black holes since the BCGs are thought to have been built up via mergers, including those involving giant elliptical galaxies, at $z \lesssim 1$.  However, given the conditions in cores of cool-core systems, we would expect that if one of the black holes is powering jets, the other ought to be too.  There appears to be no evidence of dual jets in the cores of Perseus and Virgo, nor are we aware of any additional evidence suggesting the existence of binary supermassive black holes in either Perseus, Virgo, or in any of other well-studied cool core clusters.    

 Precession can also occur if the black hole is surrounded by a geometrically-thin accretion disc whose angular momentum is misaligned with the spin axis of the black hole.   This is the more commonly invoked of the two precessions schemes \citep[e.g.][]{klein99,falceta10} to explain the observations in Perseus and Virgo.  This scenario, however, is problematic for a number of reasons, the most important of which is that the thin disc-SMBH misalignment is a relatively short-lived phenomenon with a lifetime of tens of Myrs  and over this timescale, the BH will typically {\it only undergo at the most one precession cycle}.   Both  \citet{dunn06} and \citet{falceta10} find that the Perseus observations can only be understood if the BH undergoes several precession cycles over the course of $\sim$100 Myrs. 

 As discussed in detail in \citet{lodato06} and summarized in section \ref{sec-model}, the main reason for the short lifetime is that not only does a misaligned accretion disc  induce precessional torques on the SMBH, it also  induces torques that causes the black hole spin axis to slew and move towards alignment with the total angular momentum of the black hole+disc system.    Once alignment is achieved, precession ceases.  However, the fact that  a misaligned accretion disc can cause a black hole's spin axis to tilt on timescales of tens of Myrs \citep{sch96,np98,lodato06} is intriguing and forms the basis of our proposed model.  Specifically, we assert that  while AGNs at the center of cool-core clusters typically accrete gas at a relatively low rate via radiatively inefficient, geometrically thick flows , every so often the mass accretion rate will spike and give rise to short-lived, geometrically thin accretion discs.  In general, we {\it do not} expect the angular momentum vector of these recurring discs to be aligned with the direction of the SMBH's spin.  These recurring misaligned thin discs are  central to our model because only geometrically thin structures can cause the BH to slew rapidly; geometrically thick flows do not appear to have the same effect on the SMBH \citep{kin05,fragile07,dexter11}.

\section{Titling Supermassive Black Holes}\label{sec-model}

There are a number of  questions to address in the context of our ``misaligned accretion disc'' model:  Do the existent conditions at the centers of cool-core clusters allow for the recurring formation of geometrically thin accretion discs?    Are these discs likely to be massive enough to cause the SMBH's spin axis to change direction significantly on short timescales? 

\subsection{Recurring, Short-lived High Accretion Events in Cool-Core Environments}

Apart from the hot diffuse gas,  central regions of BCGs in  cool-core clusters also appear to be threaded by a filamentary network of cold gas (\cf M87 [Virgo] \citet{forman07}; CL09 \citet[][and references therein]{osullivan2012}; NGC 4696 [Centaurus] \citet{crawford05}; NGC 1275 [Perseus] \citet{conselice01}) that generally appear to cover a significant  fraction of the $4\pi$ steradian about the cluster center.   Increasingly detailed kinematical studies of the filaments suggest that these likely have their origins in number of different physical processes, 
many of which  are also expected to induce short-lived accretion events during which the instantaneous mass accretion rate onto the central SMBH can exceed $0.01\; \dot{{\rm M}}_{\rm Edd}$,  the threshold accretion rate above which the geometrically thick, radiatively inefficient flow is expected to transition to a geometrically thin, disc flow \citep{esin97,jester05}.   Such processes include  buoyantly-rising jet-inflated bubbles  \citep{hatch06,pope10}, thermal instabilities (\citealt{piz05,sha10,mcc11,sha11}), and small-scale turbulence induced, for example, by supernova blastwaves \citep{hob11} associated with star formation in the central regions of the BCGs (\eg~\citealt{bild08}) or even by the jets themselves \citep{gaspari11}.

In the case of  buoyantly-rising jet-inflated bubbles, the bubbles are expected to be trailed by filamentary wakes of cool gas involving $\sim 10^8\; {\rm M}_\odot$ of cool gas per bubble \citep{pope10}.  In due course, this wake --- either wholly or in fragments --- will fall back towards the SMBH.    Since the filaments are primarily radial in orientation, we expect that  the fragments will have intrinsically low angular momentum (in magnitude) relative to the SMBH and have a high likelihood of ultimately settling and forming an accretion disc.  Moreover, given the geometry involved, we expect that this accretion disc to be oriented more or less perpendicular to the black hole's spin/jet axis.

As for thermal instabilities and small-scale supersonic turbulence, numerical simulation (e.g., \citealt{sha11,gas11,hob11}) studies show that both give rise to gaseous filaments, streams and high density clouds that will then ``rain'' down onto the central regions of the BCG/cluster.  In the thermal instability model, these structures are spawned during distinct cooling episodes in the cluster cores while in the case of turbulence, they are formed by convergent turbulent flows.  Only those filaments and streams with angular momentum small enough such that their circularization radius is $\sim 0.1$ pc will give rise to sub-parsec scale accretion discs.  Treating individual streams and filaments as coherent structures, simulations (e.g., \citealt{hob11}) show that the orientation of successive accretion discs that are expected to arise will be uncorrelated and can differ by large angles.

\subsection{Misaligned accretion discs and spinning black holes}\label{misalign}
 
Given an accretion disc whose  initial rotational axis is misaligned with the spin axis of the black hole, frame dragging by the rotating black hole will induce a torque on the inner regions of the disc that will  cause it to precess differentially, an effect known as Lense-Thirring precession.  \citet{bp75} showed that the viscous forces in the disc will damp the differential precession, and force the angular momenta of the disc to align\footnote{Here, we use the term ``align" loosely to refer to either co-alignment or counter-alignment.   Details of how the disc orientation evolves is immaterial to the present discussion.  What is important is that the gas is subject to torques that drives a change in its orientation.} with the total angular momentum of the system \citep{kin05,lodato06}.  The alignment of the disc proceeds from inside out, with the transition radius between the inner disc and the outer disc demarcated by a warp.   Since the process is primarily driven by frame-dragging, the influence of which falls off rapidly with distance, the warp will 
stall at a radius $R_{\rm w}$, where the rate at which the disc is twisted by Lense-Thirring precession is balanced by the rate at which viscous torques can dissipate the twist \citep{sch96,kin05,lodato06}.  This radius is given by (see Eq. 22 in \citealt{kin05})
\begin{align*}
R_{\rm w} \approx  2.7 \times  & 10^{-3} \;  j_{0.1}^{5/8}\;  \dot{{\rm M}}_{0.04}^{-1/4}\; \mbh_{9}^{9/8}\\
&\;\;\;\;\;\;\;\;\;\times \left({\alpha_2 \over 3}\right)^{-1/2}\left({\alpha_2/\alpha_1 \over 30}\right)^{-1/8}\;{\rm pc}
,\tag{2}
\end{align*}
where 
$\dot{{\rm M}}_{0.04}\equiv ({\dot{{\rm M}} / 0.04\;\dot{{\rm M}}_{\rm Edd}})$ and $j_{0.1} \equiv (j/0.1)$.  Here, $j$ is the black hole spin parameter scaled to 
$J_{\bullet,max}=(G\mbh^2/c)$, the maximum angular momentum of a Kerr black hole: $0\leq j\leq 1$.  King and collaborators have argued in a series of related articles \citep[\cf][and references therein]{kph08} that if BHs grow primarily via recurrent  randomly orientated accretion events, they will tend to have low spins.   For this reason, we have chosen to scale $j$ to a fiducial value of $0.1$.   Also,  $\alpha_1$ in the above relationship is the usual accretion disc viscosity parameter characterizing the radially outward transport of gas angular momentum and the inward transport of matter, and $\alpha_{2}$ is the viscosity associated with vertical motions in the disc.  For the purposes at hand, we  assume that over the regions of interest, the $\alpha_1$ parameter  is approximately constant and adopt $\alpha_1\approx 0.1$ as a fiducial value.   This is consistent with results from MHD simulations of magnetized accretion flows, which find that the effective value of $\alpha_1$ is  $\sim 0.1$ over the bulk of the flow (e.g., \citealt{hawley01,hawley02}).  Moreover, \citet{lodato10} show that for a strongly warped disc,  $\alpha_2 \sim 3$.  In thin discs, the warp will propagate across the disc diffusively on timescale $\alpha_2/\alpha_1$ shorter than the mass accretion timescale, which is determined solely by $\alpha_1$.    Finally, we note that we have scaled the mass accretion rate to $ 0.04\;\dot{{\rm M}}_{\rm Edd}$, which corresponds to  $\sim 1\msun {\rm yr}^{-1}$ for a $10^9 \msun$.   
  
The disc is not the only structure to experience a torque.  The black hole too will experience equal and opposite torques exerted by each differentially precessing disc annulus, which in turn will cause the black hole's spin axis to change direction.   
Discs whose outer radius is less than $R_{\rm w}$ are not expected to impact the BH in any significant fashion; consequently, we will restrict ourselves to larger disc systems.  
In such cases, the torques on the black hole is primarily due to the  gas flowing through the warp \citep{kph08}.  These torques can be resolved into two independent components: one that drives the precession of the hole's spin about the total angular momentum axis of the disc+BH system and another that drives the alignment of the black hole's spin with the total angular momentum.   \citet{lodato06} and \citet{martin07} have investigated these two processes in detail and find that the BH precession timescale is comparable to, and in realistic cases perhaps even a factor of a few longer \citep{martin07} than, the timescale over which the black hole's spin  will align with the total angular momentum of the BH+disc system: \ie, $t_{\rm BH, prec}  \gtrsim  t_{\rm align}$, where the black hole alignment timescale (Eq. 15 in \citealt{lodato06}; see also \citealt{sch96,np98}):
\begin{align*}
t_{\rm align } \approx 2 \times 10^6\;  &
j_{0.1}^{11/16}\; \dot{{\rm M}}_{0.04}^{-7/8}\;\mbh_{9}^{-1/16}\\
 &\times  \left({\alpha_2 \over 3}\right)^{1/4}\left({\alpha_2/\alpha_1 \over 30}\right)^{-15/16}\;   {\rm yrs}
.\tag{3}
\end{align*}
Consequently, the black hole is unlikely to execute more than one precession cycle, if that, and once alignment is achieved, the torques driving the precession vanish as well.    

Whether the BH attains full alignment during a given accretion event depends on the size (mass) of the accretion disc.  The disc must be sufficiently long-lived to sustain gas flow (and therefore, the torques) over the duration $t_{\rm align }$; however,
we do not necessarily require 
the BH to achieve full alignment with the total angular momentum during each and every accretion event, only that the BH's spin axis change direction.   This, as we discuss 
below, implies that only discs with  $M_{\rm d} > 10^6\;\msun$ are of interest.

At the same time, if the radially inward flow of gas in the thin-disc mode is primarily due to local viscosity,  the amount of gas that a black hole can accrete during any one accretion episode cannot be arbitrarily large  regardless of the amount of gas that is channeled into the nuclear region during the event.   The accretion discs will be susceptible to gravitational fragmentation beyond  a radius $R_{\rm frag}$ where the disc mass exceeds \citep[\cf][and references therein]{goodman03,tqm05,kp07,kph08}
\begin{equation}
M_{\rm d}(<R_{\rm frag}) =f\; \left({H\over R}\right)\mbh,\tag{4}
\end{equation}
where the factor, $f \sim$ a few, represents the uncertainty in this relationship due poorly understood details such as the extent to which the magnetic pressure, radiation pressure and stellar feedback augments thermal pressure due to viscous heating and contributes to the stability of the disc.   Adopting a simple model of \citet{collin90} to describe the properties of the sub-parsec-scale thin accretion disc (see also \citealt{kph08})\footnote{This is a model of a flat steady-state disc and we acknowledge that a flat disc and a warped disc are unlikely to have identical structures.   The use of this model is a necessary simplification in order to estimate relevant disc mass and length scales.  We note, for completeness, that  the scaling properties of this disc model are similar to that derived by \citet{ss73}.}, the total disc mass inside radius $R$, where $R$ denotes the  distance from the BH, is
\begin{align*}
M_{\rm d} \approx 3  \times & 10^6 \;  \dot{{\rm M}}_{0.04}^{3/5}\; \mbh_{9}^{4/5}\;
 \left({\alpha_1\over 0.1}\right)^{-4/5} 
R_{0.05}^{7/5}
\; \msun, \tag{5}
\end{align*} 
and the accretion disc scale height $H$ is 
\begin{align*}
\frac{H}{R} \approx  1.6\times 10^{-3}\;  \dot{{\rm M}}_{0.04}^{1/5}\;\mbh_{9}^{-3/20}\;
\left({\alpha_1\over 0.1}\right)^{-1/10}\;R_{0.05}^{1/20}. \tag{6}
\end{align*}
Here, $R_{0.05}\equiv (R/0.05\,{\rm pc})$.  Using these equations, we find that 
\begin{align*}
R_{\rm frag} \approx 0.1\;   f_5^{20/27}\;
\dot{{\rm M}}_{0.04}^{-8/27}\;  \mbh_{9}^{1/27}\;
\left({\alpha_1\over 0.1}\right)^{14/27}\;{\rm pc}. \tag{7}\\
\end{align*} 
Here, $f_5=(f/5)$.  Most of the gas at $R>R_{\rm frag}$ is expected to either turn into stars  or be expelled by stars that do form on timescales much shorter than those that govern the accretion flow.   Since only gas that flows through the warp induces a torque on the black hole \citep{kph08}, this means that the maximum amount of mass that can participate in the alignment process (derived from combining equations 5 and 7) is 
\begin{align*}
M_{\rm d,max} \approx    7.2 \times & 10^6\; f_5^{28/27}\; 
\dot{{\rm M}}_{0.04}^{5/27}\\
&\;\;\;\;\;\;\;\times \mbh_{9}^{23/27} \;
 \left({\alpha_1\over 0.1}\right)^{-2/27} \; \msun. \tag{8}
\end{align*}
We note, however, that this mass constraint can be circumvented if the mass flow through the accretion disc is mediated by global gravitational torques \citep[\cf][]{tqm05,hq11}.

Finally, we can estimate the maximum angle $\psi_{\rm max}$ through which the black hole's spin axis will slew during any one accretion event where the initial misalignment between the disc angular momentum and the black hole spin vector is characterized by angle $\theta_i$.   Following \citet{kph08}, we identity $J_{\rm d}$, the ``disc angular momentum", as the total angular momentum of the gas ($\Delta M_{\rm gas}$) flowing through the warp
during an  accretion event. That is,
\begin{equation}
J_{\rm d}= \left(G\mbh\;R_{\rm w}\right)^{1/2} \; \left(\Delta M_{\rm gas}\right).\tag{9}
\end{equation}
Straightfoward geometry then gives 
\begin{equation}
\sin\psi_{\rm max} = {J_{\rm d}\over J_{\rm h}} \sin(\theta_i-\psi_{\rm max}).
\tag{10}
\end{equation}
The above equation differs slightly from that given by \citet{kph08} because the latter's derivation implicitly assumes that $\psi_{\rm max} \ll 1$.
The extent of the tilt induced by an accretion disc depends on the initial misalignment between the disc's angular momentum and the BH's spin axis as well as on the ratio 
\begin{equation}
{J_{\rm d}\over J_{\rm h}} ={\sqrt{2}\over j}\;  \left({\Delta M_{\rm gas}\over \mbh}\right) \left({R_{\rm w}\over R_S}\right)^{1/2}.
\tag{11}
\end{equation}
where $R_S\equiv 2GM/c^2\approx 10^{-4}\;\mbh_9\; {\rm pc}$ is the Schwarzschild radius.   
To effect a tilt of $\sim 5^\circ$ in the spin axis of a slowly spinning (\ie $j\simeq 0.1$) $10^9\msun$ BH, the minimum amount of gas required is $\Delta M_{\rm gas}\simeq 10^6\msun$.    More generally, the maximum tilt (corresponding to $\theta_i=90^\circ$) experienced by a  $j\simeq 0.1$ $10^9\msun$ BH is given by
\begin{equation}
\psi_{\rm max} \approx \tan^{-1}\left( 73.5 {\Delta M_{\rm gas}\over \mbh} \right)\tag{12}
\end{equation}
For accretion discs in which the mass flow is governed by local viscosity, maximal discs (\ie $\Delta M_{\rm gas}\simeq M_{\rm d, max}$) can tilt the BH by as much as $\sim 30^\circ$.   In cases where the transfer of gas from the outer to the inner disc is mediated by global gravitational torques, the maximum possible angular displacement is only limited by the amount of mass available for accretion during any one accretion event.  A relatively modest value of $\Delta M_{\rm gas} \approx 3\times 10^7\;\msun$ is sufficient to effect a tilt of $\sim 65^\circ$.  Rapidly spinning black holes (\ie $j\simeq 0.9$) are, on the other hand, much harder to tilt.   A maximal disc can tilt a rapidly spinning ($j\simeq 0.9$) $10^9\msun$ black hole by, at the most, $\sim 7^\circ$.

\section{Discussion, Implications and Summary}\label{sec-discuss}

To recap, we have argued that supermassive black holes at the centers of cool-core clusters will, in addition to the accretion of hot diffuse gas from its surroundings, experience short-lived, recurring episodes during which the instantaneous mass accretion rate onto the central SMBH can approach or even exceed $\sim 1 \msun\; {\rm yr}^{-1}$.    We expect this phenomenon to be ubiquitous; it is a byproduct of a number of very different  and distinct physical processes expected to be operating in cores of cool-core clusters.   During such episodes, the flow in the immediate vicinity of the black hole will transition to a geometrically thin flow and establish an accretion disc.    

Current generation of simulation studies looking at the wide variety of processes at play at the centres of cool cluster cores are not detailed enough to provide a quantitative description of the spectrum of gas mass involved in individual high density accretion events,  the distribution of the time between accretion spikes, or a measure of how the angular momentum of gas varies between successive accretion events.   They do, however, suggest that the detailed nature of the accretion history of a SMBH at the center of a cool-core cluster is, in general, complex.  For instance, individual filaments may be prone to fragment into a train of clouds.  The uninterrupted accretion of a single train would result in a sequence of accretion events occurring in quick succession on a timescale of order the free-fall time, \ie, $\sim 10$ Myrs, and give rise to a succession of discs whose orientations vary only by small angles.   Depending on the masses of individual discs and the time between successive events, the SMBH's spin could tilt through several discrete but correlated small angular displacements or one seemingly large displacement.   Still, if one treats the accretion of an individual filament, stream or a train of associated clouds as a single coherent event, the simulations do suggest that the direction of the spin axis of the SMBH will vary stochastically from event to event and over the course of many such events, the orientation of the black hole's spin axis will execute a random-walk over $4\pi$ steradian.  

Within the context of the turbulence model, we can attempt to estimate the mean time between accretion events.   Turbulence is typically expected to give rise to streams and clouds with mass spectrum \citep{hqm12}
\begin{equation}
dN_{\rm cl} \propto M_{\rm cl}^{-1.8}\;\; dM_{\rm cl},\;\;\;\;\; M_{\rm cl} \lesssim f^2_{\rm gas}M_{\rm gas}(<R),\tag{13}
\end{equation}
where $f_{\rm gas}$ is the fraction of gas mass relative to the {\it total} enclosed mass.   The mass of cold gas  in cool-core cluster BCGs is $\sim$ a few $\times10^{10}\;\msun$ within the central 10 kpc \citep{edge01,sc03,edge10}, which corresponds to $f_{\rm gas}\sim 0.1$.   In this case, the number of clouds with mass $M_{\rm cl} \gtrsim 10^{6}\msun$ is $\sim$2,500.  (We choose this mass threshold because, as discussed in \S 3.2, less massive clouds have minimal impact on the SMBH.)   Only a fraction of these clouds, with angular momentum small enough such that the circularization radius is $\sim 0.1$ pc, will give rise to accretion discs.   Assuming that the velocity distribution of the clouds is isotropic, with velocity dispersion $\sigma_{\rm cl}\approx 300$ km/s, and that the radial distribution of the clouds is approximately isothermal ($n_{\rm cl}\propto r^{-2}$), the mean time between accretion events is $\sim$ a few $\times 10^7$ yrs.  This timescale is in agreement with that implied by the history of AGN activity in Perseus and Virgo, as summarized in Table 1:  Perseus and Virgo data both suggest $\sim 5 $ events over the past $10^8$ years.

When an accretion event gives rise to a disc whose orientation is misaligned with the spin axis of the black hole, 
the resulting torques between the gas flowing through the disc and the black hole will cause the latter's axis to swivel and change direction.   The magnitude of the change in the direction of the BH's spin axis depends on the degree of the initial misalignment between the disc's angular momentum and the BH's spin axis, the magnitude of the SMBH's spin, and the amount of gas that flows through the disc and accretes onto the BH.   If the inward flow of gas is mediated by local viscous stresses, we find that a misaligned disc can cause the spin axis of a slowly rotating (\ie, $j=0.1$) SMBH to slew by as much as $\sim 30^\circ$.  If, however, the inward flow is induced by global gravitational torques, the BH can potentially tilt by much larger angles.

Drawing on the considerable body of work indicating that jet production is most efficient when the accretion flow is geometrically thick and suppressed otherwise \citep{livio99,meier01} \citep[\cf also][and references therein]{bb09}, we expect that the jets will briefly wane in power during the thin disc phase.\footnote{This statement should not be interpreted as our suggesting that  jet activity cannot arise during the quasar phase.   We appreciate that at sufficiently high mass accretion rate, the flow should be able to sustain both behaviours.   However, we do not expect such configurations to be common in cluster environments.}  
But once the accretion disc drains away and the geometrically thick flow re-establishes, the jets will also resume.   Since the jet axis is the same as the spin axis, the post-tilt jets will point in a different direction from the pre-tilt jets.   This scenario offers the simplest explanation of, for example, the apparently abrupt change in direction between an older (larger) relic jet in CL09 and a subsequent jet episode \citep{osullivan2012}.

The scenario outlined in this paper and summarized above has several important astrophysical implications:

\smallskip
\noindent 
(1) The recurring reorientation of AGN jets due to tilting of the SMBH spin axis offers a straightforward explanation for the distinct, randomly oriented jets/lobes/cavities observed in cool-core clusters such as Perseus, Virgo and CL09.

\smallskip
\noindent 
(2)  Swivelling jets also offer a simple yet compelling resolution for the ``isotropic heating'' puzzle.  Since the jets are expected to change directions several times over the cooling time in cores of cool-core clusters, the resultant heating will also be distributed over a large portion of $4\pi$ steradians on the same timescale.     

\smallskip
\noindent 
(3) Since geometrically thin accretion discs are radiatively efficient, our model predicts that whenever one forms, the host AGN ought to transform into a quasar.  The quasar will be short-lived because the discs do not involve a lot of  mass and are expected to drain quickly.   We can estimate the likelihood of catching an AGN ``in the act" as follows:
The Perseus and Virgo data both suggest $\sim 5$ events over the past $10^8$ yrs.   Assuming that a typical accretion event involves  $\sim 10^6\msun$ of gas and that the lifetime of the resulting disc is $\sim$ Myrs,  we expect a quasar duty cycle of $\sim 5\%$.    

\smallskip
\noindent 
(4) A $\sim 5\%$ duty cycle means that out of $\sim 250$ rich clusters ($T_x > 2$ keV) with $z < 0.5$ that have X-ray observations, we ought to expect  
1--2 clusters to host a central quasar.   This estimate is based on the following:  Approximately $35\%$ of rich clusters \citep{emp11} tend to be strong cool core systems and of these, only 35\% show evidence of significant multiphase gas component in their cores \citep{macdonald10}.   Only the AGNs in the latter systems are likely to experience enhanced accretion events on a recurring basis.  Interestingly, there are in fact only two known $z<0.5$ clusters that host quasars:  MACS J0913.7+4056 [also known as CL09104+4109] \citep{osullivan2012} at $z=0.44$ which hosts a dust enshrouded type 2 QSO at its centre, and a $z= 0.3$ cluster that hosts a highly luminous radio-quiet quasar, H1821+643 \citep{rus10}.   Both quasars are located at the centres of cool-core clusters.

\smallskip
\noindent 
(5) Finally,  we note that our proposal --- that SMBHs at the centres of cool-core clusters are repeatedly tilted by misaligned accretion discs --- implicitly requires the SMBHs {\it in such environments} to be relatively slow rotators, which they are likely to be if they have accreted a non-negligible fraction of their mass via randomly aligned streams, filaments and clouds.  If they are shown to be spinning rapidly, our proposed mechanism cannot explain the observations.  Thin misaligned discs cannot tilt rapidly spinning SMBHs by more than a few degrees.

\section{Acknowledgments}
The authors acknowledge the hospitality of the Kavli Institute for Theoretical Physics at the University of California Santa Barbara where this work was conceived.    
This research was supported in part by the National Science Foundation under Grant No. NSF PHY05-51164 to KITP.   AB acknowledges support from NSERC Canada through the Discovery Grant program and CSR acknowledges support from the National Science Foundation under Grant No. AST0908212.   The authors would also like to thank P. Ajith, A. Benson, F. Durier, N. Murray, and G. Novak for useful discussions.
 
\vfill\eject

\newpage

\end{document}